%% file: main.tex
\pdfoutput=1
\documentclass[10pt,conference]{IEEEtran}
\IEEEoverridecommandlockouts
\usepackage{cite}
\usepackage{amsmath,amssymb,amsfonts}
\usepackage{algorithmic}
\usepackage{graphicx}
\usepackage{textcomp}
\usepackage{xcolor}

\usepackage{amsmath,amsfonts}
\usepackage{physics}
\usepackage{bm}

\usepackage{array}
\usepackage{multirow}
\usepackage{tabularx}
\usepackage{graphicx}
\usepackage{caption3}
\usepackage{color, colortbl}
\usepackage{tcolorbox}
\usepackage{siunitx}
\usepackage{subcaption}
\usepackage{listings}
\usepackage{float}
\usepackage{makecell}
\usepackage{tabularx}
\usepackage[flushleft]{threeparttable}

\usepackage[ruled,linesnumbered,vlined,noend]{algorithm2e}

\SetCommentSty{mycommfont}

\usepackage{amsthm}

\theoremstyle{definition}

\newtheorem*{definition*}{Definition}

\definecolor{dkgreen}{rgb}{0,0.6,0}
\definecolor{gray}{rgb}{0.5,0.5,0.5}
\definecolor{mauve}{rgb}{0.58,0,0.82}
\definecolor{codegreen}{rgb}{0,0.6,0}
\definecolor{codegray}{rgb}{0.5,0.5,0.5}
\definecolor{codepurple}{rgb}{0.58,0,0.82}
\definecolor{saddlebrown}{rgb}{0.55,0.27,0.075}
\definecolor{backcolour}{rgb}{0.9,0.9,0.9}
\lstdefinestyle{base}{
  language=C,
  emptylines=1,
  breaklines=true,
  basicstyle=\ttfamily\footnotesize
}

\newcolumntype{M}[1]{>{\centering\arraybackslash}m{#1}}

\def\BibTeX{{\rm B\kern-.05em{\sc i\kern-.025em b}\kern-.08em
    T\kern-.1667em\lower.7ex\hbox{E}\kern-.125emX}}

\begin{document}

\title{\textsc{Iota}: A Framework for Analyzing System-Level Security of IoTs
}

\author{
    \IEEEauthorblockN{
        Zheng Fang\IEEEauthorrefmark{1},
        Hao Fu\IEEEauthorrefmark{1}, 
        Tianbo Gu\IEEEauthorrefmark{1},
        Pengfei Hu\IEEEauthorrefmark{2},
        Jinyue Song\IEEEauthorrefmark{1}
        Trent Jaeger\IEEEauthorrefmark{3}, and
        Prasant Mohapatra\IEEEauthorrefmark{1}
    }
    \IEEEauthorblockA{
        \IEEEauthorrefmark{1}
        Department of Computer Science, University of California, Davis\\
        \IEEEauthorrefmark{2}
        School of Computer Science and Technology, Shandong University\\
        \IEEEauthorrefmark{3}
        Department of Computer Science and Engineering, Pennsylvania State University\\
        Email: \{zkfang,haofu,tbgu,jysong,pmohapatra\}@ucdavis.edu, phu@sdu.edu.cn, tjaeger@cse.psu.edu
    }
}

\maketitle

\begin{abstract}
Most IoT systems involve IoT devices, communication protocols, remote cloud, IoT applications, mobile apps, and the physical environment. However, existing IoT security analyses only focus on a subset of all the essential components, such as device firmware, and ignore IoT systems' interactive nature, resulting in limited attack detection capabilities. In this work, we propose \textsc{Iota}, a logic programming-based framework to perform system-level security analysis for IoT systems. \textsc{Iota} generates attack graphs for IoT systems, showing all of the system resources that can be compromised and enumerating potential attack traces. In building \textsc{Iota}, we design novel techniques to scan IoT systems for individual vulnerabilities and further create generic exploit models for IoT vulnerabilities. We also identify and model physical dependencies between different devices as they are unique to IoT systems and are employed by adversaries to launch complicated attacks. In addition, we utilize NLP techniques to extract IoT app semantics based on app descriptions. To evaluate vulnerabilities' system-wide impact, we propose two metrics based on the attack graph, which provide guidance on fortifying IoT systems. Evaluation on 127 IoT CVEs (Common Vulnerabilities and Exposures) shows that \textsc{Iota}'s exploit modeling module achieves over 80\% accuracy in predicting vulnerabilities' preconditions and effects. We apply \textsc{Iota} to 37 synthetic smart home IoT systems based on real-world IoT apps and devices. Experimental results show that our framework is effective and highly efficient. Among 27 shortest attack traces revealed by the attack graphs, 62.8\% are not anticipated by the system administrator. It only takes 1.2 seconds to generate and analyze the attack graph for an IoT system consisting of 50 devices. 
\end{abstract}


\input{sections/introduction}
\input{sections/threat_model}
\input{sections/system_overview}
\input{sections/design}
\input{sections/metrics}
\input{sections/implementation}
\input{sections/evaluation}
\input{sections/future_work}
\input{sections/related_work}
\input{sections/conclusion}
\input{sections/acknowledgement}

\bibliographystyle{IEEEtranS}
\bibliography{main.bbl}

\end{document}

%% file: sections/introduction.tex
\section{Introduction}
The last decade witnessed the rapid development and wide deployment of IoT systems. According to \cite{intel2020}, the total global worth of IoT technology could be as much as 6.2 trillion US dollars by 2025. Popular commodity IoT platforms, such as Samsung SmartThings \cite{platform:smartthings}, Apple Homekit \cite{platform:homekit}, and Google Nest \cite{platform:googleSmartHome}, etc., share similar architecture: low power end devices running customized firmware, short-range, wireless communication protocols, a centralized decision-maker, IoT applications using trigger-action paradigms, and companion mobile apps. IoT components interact with each other in sophisticated ways. For example, devices' functionality depends on secure and reliable communication with the controller, and devices can be dependent on each other due to IoT applications or physical channels. The distributed but interactive components pose tremendous challenges to IoT system security verification and analysis \cite{nguyen2018iotsan, celik2018soteria, ding2018safety, wang2019charting, fang2019foresee}. 


Existing research on IoT security only focuses on a single or a subset of the IoT components. For instance, \cite{costin2014large, zheng2019firm, cui2013firmware} analyze IoT device firmware, \cite{ronen2017iot} investigates IoT wireless protocols, and \cite{nguyen2018iotsan, celik2018soteria, trimananda2020understanding} sanitize IoT applications. However, for interconnected systems, hardening individual components cannot guarantee security because there are multiple paths to compromise system resources. For example, attackers can unlock a smart doorlock by exploiting vulnerabilities on the lock \cite{cve:yale}, but they may also compromise an indoor camera \cite{cve:nestcam} and use it to inject voice, triggering a smart speaker to launch the door-open command \cite{attack:alexa}. In this paper, we try to address the following research problem --- How to verify IoT systems security and uncover threats in a systematic way? 


Attack graphs \cite{ou2006scalable, sheyner2002automated, ammann2002scalable} provide us an elegant approach to the problem by enumerating all of the paths to potential \emph{attack goals}, i.e., system resources which can be compromised by the attacker.
There are two types of attack graphs: \emph{state-based attack graph} \cite{ritchey2000using, sheyner2002automated} and \emph{exploit-dependency attack graph} \cite{ammann2002scalable, ou2006scalable}. State-based attack graphs utilize model checking as the reasoning engine. But they suffer scalability issues in that the size of the graph grows exponentially with the number of system state variables (The number of system state variables is a linear function of the system device count). In comparison, it takes polynomial time to construct exploit-dependency attack graphs, and the generated attack graph size is a quadratic function of the system device count \cite{ou2006scalable}.

However, existing exploit-dependency attack graph frameworks cannot be readily applied to IoT systems due to multiple design limitations. First, existing exploit-dependency attack graphs were designed for conventional computer networks and do not model essential IoT components such as IoT apps and devices' physical dependencies. Second, many IoT devices communicate using low-power protocols such as Zigbee or ZWave, which most of the existing vulnerability scanners cannot scan. For example, all of the vulnerability scanners listed on \cite{owasp:scanner} only support IP-based devices. Moreover, existing frameworks do not model exploits on low-power, short-range protocols which are ubiquitous in IoT systems. Finally, there are no quantitative criteria for administrators to harden the system in such a way that vulnerabilities with the largest impacts get patched first. As today's IoT systems may contain hundreds of vulnerabilities, it is necessary to patch vulnerabilities efficiently. 

\textbf{Goals}. In this paper, our goal is to build a system-level security analysis framework for IoTs which, given the IoT system configurations (i.e., device, network information, and the IoT apps installed), (a) constructs exploit-dependency attack graphs to uncover resources that can be compromised and reveal potential attack traces; and, (b) computes a suite of metrics to interpret the generated attack graph and provide recommendations for system hardening.

As exploits and devices' dependencies are the key building blocks of attack graphs, to achieve (a), we extract exploit models and device dependencies from IoT system configurations and represent them as Prolog clauses \cite{ou2005mulval}. More specifically, \textsc{Iota} scans IoT system configurations for individual vulnerabilities and builds \emph{exploit models} (consisting of precondition and effect) based on scanned CVEs. We identify three types of device dependencies: \emph{app-based dependency}, \emph{indirect physical dependency}, and \emph{direct physical dependency}. The app-based dependencies are specified by IoT app semantics (i.e., trigger-action rules). Since IoT apps' source code can be unavailable in some platforms, such as IFTTT \cite{platform:ifttt}, we utilize natural language processing (NLP) techniques to extract app semantics from app descriptions. The direct and indirect physical dependencies are universal in IoT systems and thus are hard-coded as Prolog rules. Finally, Prolog clauses are sent to MulVAL \cite{ou2005mulval} to generate attack graphs.

With regards to (b), we propose two novel metrics: \emph{shortest attack trace} to an attack goal, and \emph{blast radius} of a vulnerability. The shortest attack trace to an attack goal node provides the lower bound of the attack complexity in terms of the number of exploits to launch. A vulnerability's blast radius tells us the potential capabilities the attacker can get on the system by exploiting \emph{only} that vulnerability.
In addition, the concept of \emph{attack evidence} (defined to help us compute blast radius) can also be used to compute the \emph{minimal set of vulnerabilities to patch to thwart an attack goal} \cite{sheyner2002automated}. These metrics help administrators interpret the attack graphs, sort the vulnerabilities based on their impacts on the system, and make informed choices about system hardening.

To evaluate \textsc{Iota}, we generate 37 synthetic smart home IoT systems based on 532 real-world IoT apps and a list of 59 smart home devices of 26 types. We scan the CVE database since 2010 and find 127 IoT CVEs on those 59 smart home devices. Our vulnerability analysis module achieves $80.56\%$ accuracy for exploit precondition identification and $88.19\%$ accuracy for exploit effect. We manually check 27 shortest attack traces whose depths are at least 9 and find out 62.8\% of them are beyond anticipation. In particular, the graph analyzer module reveals a shortest trace of depth 18 for an IoT system consisting of only 7 devices, implying that attack traces can be very deep for even a small IoT system. The case study illustrates the effectiveness of using the shortest attack trace and blast radius to estimate attack complexity and their impacts on the system. \textsc{Iota} is highly scalable. In practice, it only takes around 1.2s and 120MB memory to evaluate IoT systems of 50 devices. 

In summary, we make the following contributions:
\begin{itemize}
    \item We introduce \textsc{Iota}, a novel framework to conduct automatic, system-level IoT system security analysis and generate attack graphs showing all potential attack traces.
    \item We design formal models for IoT exploits and implement automatic translation from system configuration and vulnerability information to Prolog clauses.
    \item We propose two metrics to quantitatively evaluate the attack complexity (shortest attack trace) and vulnerability's system-level impacts (blast radius).
    \item We evaluate the efficiency and effectiveness of \textsc{Iota} by applying it to 37 synthetic IoT systems of different sizes ranging from 4 to 50 and verify that our framework is both effective and highly efficient.
\end{itemize}


%% file: sections/threat_model.tex
\section{Threat Model}\label{sec:threat}
In this work, we consider individual attackers whose goal is to break into the system. They can be physically adjacent to the IoT system, enabling them to be within the radio range of the wireless local area networks, such as WiFi or Zigbee networks. Besides, the attacker can physically access outside, unprotected IoT devices, such as a doorbell or outdoor surveillance cameras. We also assume the attacker is able to extract IoT app semantics because he can install sniffers and infer event type from the sniffed packets \cite{gu2020iotgaze}. We treat the remote IoT cloud as trustworthy and do not consider the compromise of the cloud itself. However, if there exist vulnerabilities on the companion mobile app, the attackers can spoof commands to the remote cloud. Below we discuss the major threats to each of the IoT components in detail.

\begin{figure*}[ht]
\centering
\includegraphics[scale=0.47]{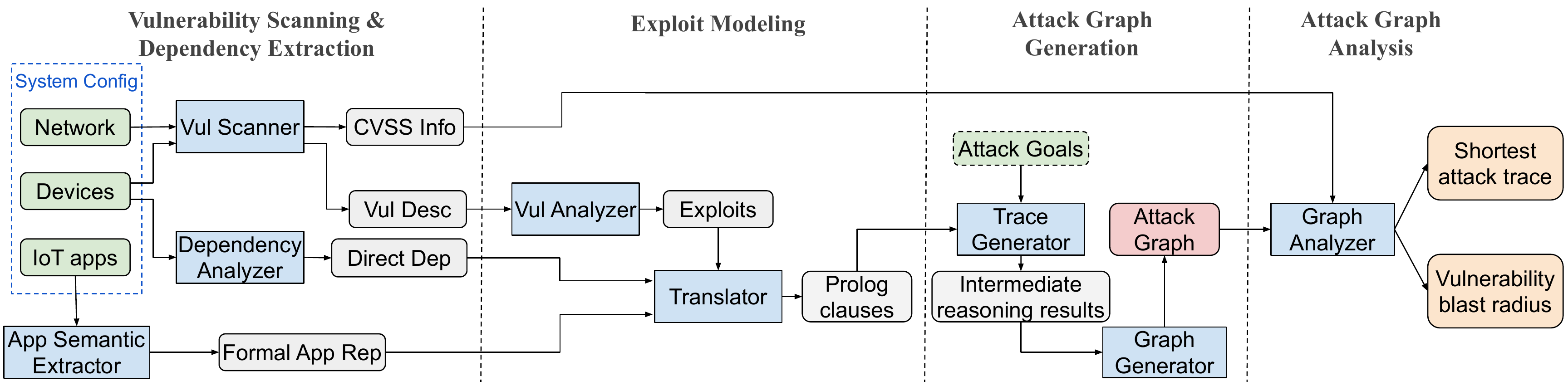}
\caption{\protect\textsc{Iota} pipeline. The blue boxes are \textsc{Iota} modules. The green, red, and gray boxes represent input, output, and intermediate results, respectively. The \emph{Attack Goals} can be set by the system administrator and is optional.}
\label{fig:flow}
\vspace{-4mm}
\end{figure*}



\subsection{Device}
We use the term ``IoT devices'' to refer to both end devices and infrastructure devices such as routers and gateways. Most of the device vulnerabilities are rooted in the firmware 
\cite{costin2014large, costin2016automated, hernandez2017firmusb}. However, some vulnerabilities are found in the device's physical components \cite{son2015rocking, sugawara2020light}. Once a device is compromised, it can be used to attack other components of the IoT system in three different ways. First, if the attacker gets root privilege on a device such as a router, he can send spoofed commands to other devices on the same network. Second, the attacker can utilize the compromised device to inject cyber events, such as spoofing a motion event. Moreover, the attacker can take advantage of the devices' physical dependencies to compromise other devices. 

\subsection{Network}
IoT systems utilize short-range, low-power protocols to communicate with the end devices, which allows adjacent attackers to sniff the wireless traffic. These end devices are first connected to a gateway (sometimes called base station, bridge, or hub) in order to communicate with the internet. Since generally there is no firewall or MAC address filtering in most smart home networks, if the attackers gain access to the network, they can send spoofed packets to other devices on the same network. To make things worse, many IoT devices, such as routers, cameras, or thermostats, expose unprotected network services to the home network, making it possible for the attackers to compromise these devices after they join the home network. 


\subsection{IoT application}
IoT applications are designed using the \emph{if-this-then-that} paradigm, where \emph{this} represents IoT event(s) and \emph{that} represents some device actions. IoT apps introduce dependencies among devices, which expose new attack surfaces to the adversary. Consider ``If smoke is detected, sound the alarm and open the window.'' as an example. To open the victim's window, the attacker does not have to attack the window opener directly; instead, he can just compromise the smoke detector, and the IoT app will do the rest of the attack for him. Even though the attacker must know the app has been installed before exploiting it, people have shown this can be done via wireless sniffing \cite{gu2020iotgaze, zhang2018homonit}.


\subsection{Physical channel}
One of the unique features of IoT systems compared with other networked systems is that IoT devices can interact with each other via the \emph{physical channels}. There is a distinction between the IoT physical channels and the physical layer of the computer networks: The former is shared physical environments, such as air, temperature, and humidity, whereas the latter is electromagnetic signals transmitting raw bitstreams. While IoT app-based device dependencies will only exist if the app is installed by the user, physical device dependencies \emph{always} exist in an IoT system as long as the relevant devices are installed. An attacker can utilize various physical dependencies to launch attacks. For instance, he can first compromise the indoor camera, e.g., Nest Cam IQ Indoor, and use it to inject human voice commands. The smart speaker will receive the voice and issue the corresponding command to the actuator. 

%% file: sections/system_overview.tex
\section{System Overview} \label{sec:system_design}


Figure \ref{fig:flow} illustrates the pipeline of \textsc{Iota}, which consists of four stages. \textbf{Vulnerability Scanning and Dependency Extraction} stage scans the devices and network protocols for vulnerabilities and extracts IoT app semantics. It also extracts direct device dependencies from the system configuration file. \textbf{Exploit Modeling} stage maps vulnerabilities to exploits based on the vulnerability description and Common Vulnerability Scoring System (CVSS) \cite{cvss} scores such as Attack Vector and Confidentiality, Integrity, Availability (CIA) triad. Exploits, direct dependencies, and app-based device dependencies are then translated to Prolog clauses. \textbf{Attack Graph Generation} stage reads attack goals (optional) and the translated Prolog clauses and generates IoT attack graphs. If the user does not provide attack goals, we enumerate all system resources (i.e., privileges on devices or tamper of the physical features) as potential attack goals. Then we modify the intermediate reasoning results and send them to MulVAL \cite{ou2005mulval} to generate attack graphs. \textbf{Attack Graph Analysis} stage takes the generated attack graph as input and computes the following metrics: the shortest attack traces to each attack goal node and the blast radius of each vulnerability.

%% file: sections/design.tex
\section{Iota Design} \label{sec:design}
In this section, we explain the design of the \textsc{Iota} modules as shown in Figure \ref{fig:flow}. The implementation details are explained in Section \ref{sec:implementation}.

\subsection{Vulnerability Scanner} \label{sec:vul_scanner}
To the best of our knowledge, there are no existing vulnerability scanners readily available for low power communication protocols such as Zigbee or Bluetooth Low Energy (BLE). Therefore, we design a vulnerability scanning approach based on CVE database searching. Our approach is practical because of some device vendors' ignorance of vulnerability report \cite{philips, radio:thermostat} and the slow firmware upgrade rate \cite{cui2013firmware}. 

Given the IoT devices installed and the communication protocols used, the Vulnerability Scanner module searches the CVE database \cite{cve} for vulnerabilities. We fetch the CVE JSON feeds since 2010 from the National Vulnerability Database (NVD) \cite{nvd} and parse the JSON files to get information relevant to our exploit modeling, including impact score, exploitability score, exploit range, exploit result (CIA triad), and the vulnerability description. The parsing results are stored in a local MySQL database. In total, there are 121,210 CVEs from 2010 to April 2021. After discarding CVEs without CVSS information, our database contains 113,180 records. For each device instance listed in the system configuration file, we query the database for the device name using full-text search in boolean mode to make sure it only returns CVE records when all of the query keywords appear in the CVE description.

The scanned vulnerability on each device is then translated to Prolog facts. For example, the following fact shown in Listing \ref{list:vul_exists} means vulnerability \texttt{CVE-2020-8864} exists on \texttt{dLinkRouter}.
\begin{lstlisting}[caption={Prolog fact for a CVE found on a device.},
label={list:vul_exists},]
vulExists(dLinkRouter, 'CVE-2020-8864').
\end{lstlisting}

\subsection{Dependency Analyzer} 
The Dependency Analyzer module models how IoT devices interact with each other via physical channels. We identify and define two types of physical dependencies: direct dependency and indirect dependency. Two devices are \emph{directly dependent} if they are both actuators. There are three types of direct dependencies as listed in Table \ref{table:direct_dep}. The most common direct dependency is \textbf{electrical dependency}, such as the one between smart outlet and air conditioner. The second type is \textbf{mechanical dependency}. For example, the door opener cannot open the door if the door lock is locked. We define the third type as \textbf{utility dependency}. For example, gas valve and smart stove are dependent via gas. Even though direct dependencies can have a huge impact on IoT system security, they are overlooked by existing IoT security analysis frameworks. 


\begin{table}[t]
\small
\centering
\caption{IoT device direct dependencies and examples.}
\begin{tabularx}{\linewidth}{|>{\hsize=0.14\hsize}c|>{\hsize=0.82\hsize}X|}
\hline
\rowcolor[HTML]{C0C0C0}
\textbf{Direct Dependency} & \multicolumn{1}{c|}{\textbf{Example}}  \\ \hline
\textit{Electrical} & Outlet $\rightarrow$ AC; Switch $\rightarrow$ Light bulb  \\ \hline
\textit{Mechanical} & Door lock $\rightarrow$ Door opener  \\ \hline
\textit{Utility} & Water valve $\rightarrow$ Sprinkler; Gas valve $\rightarrow$ Stove \\ \hline
\end{tabularx}
\label{table:direct_dep}
\vspace{-4mm}
\end{table}

Two devices are \emph{indirectly dependent} if one of them is an actuator and the other is a sensor. We consider and model six physical channels: temperature, humidity, illuminance, voice, smoke, and water. We include ``voice'' as a physical channel because many devices like cameras and TVs can play human voice in the smart home, and some devices can recognize human voice and execute the corresponding instructions. 

Direct and indirect physical dependencies are hard-coded as Prolog rules because they are universal in all IoT systems, regardless of the installation of certain IoT apps. During the execution of a Prolog program, a certain dependency rule will be activated only when the corresponding device is installed. For example, if AC is on, then the room temperature will be low. But if there is no temperature sensor installed, the predicate of sensor reporting low temperature will not hold true. Listing \ref{list:dev_indirect_dep} and Listing \ref{list:dev_direct_dep} are example of Prolog rules for indirect and direct dependencies, respectively.

\begin{lstlisting}[caption={Indirect physical dependency.},
label={list:dev_indirect_dep},]
high(temperature) :-
    on(Heater),
    heater(Heater).

reportsHigh(TemperatureSensor, temperature) :-
    high(temperature),
    temperatureSensor(TemperatureSensor).
\end{lstlisting}

\begin{lstlisting}[caption={Direct physical dependency.},
label={list:dev_direct_dep},]
off(Device) :-
    plugInto(Device, Outlet),
    outlet(Outlet),
    off(Outlet).
\end{lstlisting}

\subsection{App Semantic Extractor}\label{sec:app_logic_extractor}
The App Semantic Extractor module extracts semantic information from IoT app descriptions using NLP techniques. Compared with program analysis, analyzing IoT app descriptions in NLP is more applicable in that app descriptions are publicly available while IoT apps' code may be proprietary on some platforms. In smart home platforms, developers write a short description to explain the functionality of their IoT apps to smart home users. Typically, these app descriptions are written in ``If this, then that'' form, which makes it suitable for NLP techniques. 

We use Stanford CoreNLP framework \cite{manning-EtAl:2014:P14-5} and Natural Language Toolkit (NLTK) \cite{bird2009natural} for app description analysis. Given an app description, we use CoreNLP parser to construct the constituency parse tree and split the sentence into the conditional clause and the main clause based on tree node labeled \texttt{SBAR} (subordinate clause). We do a breadth-first search on the parse tree to find the tree node with label \texttt{SBAR}, which is the root of the conditional clause. Then the conditional clause is obtained by concatenating the leaf nodes of the subtree whose root node has label \texttt{SBAR}. We construct the main clause by removing the conditional clause string from the whole sentence. 

Because the conditional clause and the main clause may contain multiple conditions or actions, we further split each clause into simple sentences based on tree node labeled \texttt{CC} (coordinating conjunction). The coordinating conjunction represents either logic AND or logic OR relationship between the two simple sentences. For example, the split of SmartApp \textit{Hall Light: Welcome Home}'s description ``Turn on hall light if someone comes home and the door opens.'' is shown in Listing \ref{list:split}. The conditional clause is split into two simple sentences with logic AND relationship. Since the main clause contains just one simple sentence, the relationship is set to \texttt{'NONE'}. 

\begin{lstlisting}[
    caption={Splitting clauses into simple sentences for the SmartApp \textit{Hall Light: Welcome Home}.},
    label={list:split},]
conditional:  ('AND', ['someone comes home', 'the door opens'])
main:  ('NONE', ['Turn on the hall light'])
\end{lstlisting}



After splitting each clause into simple sentences, we extract noun and verb phrases from each simple sentence and match them to IoT device names and device action using Word2Vec similarity. We use a regular expression chunker to extract noun phrases and verb phrases. The regular expression patterns we use for chunking and the extracted phrases for the SmartApp description are shown in Listing \ref{list:regex} and Listing \ref{list:chunk}, respectively.

\begin{lstlisting}[
    caption={Regular expression patterns.},
    label={list:regex},]
NP: {<DT>?<JJ>*<NN.*>+}
VP: {<VB.*><IN|RP>?}
\end{lstlisting}

\begin{lstlisting}[
    caption={Noun and verb phrases extracted for the SmartApp \textit{Hall Light: Welcome Home}'s description.},
    label={list:chunk},]
conditional clause: [(['someone'], ['comes']), (['the door'], ['opens'])]
main clause: [(['the hall light'], ['Turn on'])]
\end{lstlisting}

Finally, we use Word2Vec model \cite{mikolov2013distributed} to match the extracted noun phrases and verb phrases with our pre-defined list of devices and device actions. Since Word2Vec only computes similarities between individual words, we compare each word in a noun phrase against each word in a device name. The app semantic extraction result is represented as a Python tuple shown in Listing \ref{list:app_logic_tuple}. This internal representation is used together with app configuration information in the Translator module to generate Prolog rules. 

\begin{lstlisting}[
    caption={Internal representation of the IoT app semantic.},
    label={list:app_logic_tuple},]
('AND', ['motion sensor', 'door contact sensor'], ['motion', 'open'], 'NONE', ['bulb'], ['on'])
\end{lstlisting}

\subsection{Vulnerability Analyzer} \label{sec:vul_analyzer}
The Vulnerability Analyzer module maps vulnerabilities to exploit models. Exploit modeling is essential for attack graph construction because attack traces are composed of individual exploits. To the best of our knowledge, our work is the first to attempt to \emph{automatically} generate exploit models based on CVEs' natural language description and CVSS scores. Though MulVAL \cite{ou2005mulval} formally represents exploits as Prolog rules, it only considers privilege-escalation in computer networks. Our exploit models are designed for generic IoT systems and consist of exploit precondition and effect. A \emph{precondition} is the privilege the attacker should have in order to launch an exploit. An \emph{effect} is the direct result of an exploit. 

\textbf{Precondition}. For IoT systems, we define five types of preconditions listed in Table \ref{table:exp_precondition}. Because IoT systems typically involve low-power, short-range, wireless protocols such as Wifi or ZigBee, the physically or logically adjacent precondition should be defined for each network type specifically, such as \texttt{Wifi adjacent logically}, \texttt{Zigbee adjacent physically}, etc. We cannot just use the ``attack vector'' value as the precondition of each CVE in the NVD database because that field can be ambiguous or sometimes incorrect: According to \cite{cvss:rubrics}, it assigns ``network'' as the precondition whenever there is a lack of information to decide the exploit range. Besides, the value does not differentiate ``physically adjacent'' and ``logically adjacent''. 

We predict the exploit precondition based on protocol type, CVE description, and the CVSS attack vector. If an exploit's attack vector is \texttt{local} or \texttt{physical}, we keep its value. If the attack vector is \texttt{adjacent}, we check its CVE description. If the description contains keywords such as ``sniff'', ``decrypt'' or their synonyms, we will assign the precondition as \texttt{adjacent physically}, otherwise \texttt{adjacent logically}. If the original attack vector is \texttt{network}, we will first check the protocol. If the protocol is a low-power protocol, then we invoke the approach of determining \texttt{adjacent}; otherwise, we set the precondition to \texttt{network}.

\begin{table}[t]
\small
\centering
\caption{Types of exploit preconditions on IoT devices.}
\begin{tabularx}{\linewidth}{|>{\hsize=0.25\hsize}c|>{\hsize=0.75\hsize}X|}
\hline
\rowcolor[HTML]{C0C0C0}
\textbf{Precondition} & \makecell[c]{\textbf{Explanation}} \\ \hline
Network    &   {An attacker can exploit the vulnerability from the internet. There's no prior privilege required on the IoT system.}          \\ \hline
Adjacent physically       &     The attacker needs to be within the radio range of a wireless network, but he/she does not need to be on the network.            \\ \hline
Adjacent logically          &      The attacker should be both within the radio range and on the wireless network, in order to launch the exploit.         \\ \hline
Local      &     The exploit requires access to the device with at least user privilege, such as establishing Telnet or SSH connection to the device.     \\ \hline
Physical      &     The attacker needs physical access to the vulnerable device to commit exploits.   \\ \hline
\end{tabularx}
\label{table:exp_precondition}
\end{table}

\textbf{Effect}. From an attacker's perspective, exploit effects include gaining privileges on IoT devices, accessing wireless traffic, or making devices denial-of-service. We categorize exploit effects into six types as listed in Table \ref{table:exp_effect}. If attackers get \texttt{root} privilege on a device, they can use it to attack other devices by sniffing or spoofing wireless traffic. The \texttt{device control} privilege implies both \texttt{command injection} and \texttt{event access} privilege, but not the capability of attacking other devices on the same system.

\begin{table}[t]
\small
\centering
\caption{Types of exploit effects on IoT devices.}
\begin{tabularx}{\linewidth}{|>{\hsize=0.245\hsize}c|>{\hsize=0.755\hsize}X|}
\hline
\rowcolor[HTML]{C0C0C0}
\textbf{Effect} & \multicolumn{1}{c|}{\textbf{Explanation}}  \\ \hline
Root    &   Attackers have root privilege on a device, which can be used to send spoofed commands to other IoT devices on the same network.  \\ \hline
Device control   &   The attacker can run any command the device supports, and sniff and inject any device event. But the device cannot be used to send spoofed messages to other devices.   \\ \hline
Command injection    &   The attacker can inject any commands to the device, but does not have access to the IoT events on that device.  \\ \hline
Event access      &    The attacker is able to sniff and spoof events on an IoT device, but he/she cannot inject commands to that device.    \\ \hline
Wifi access    &    The attacker obtains the Wifi credentials and is able to join the Wifi network. \\ \hline
DoS     &    The IoT device becomes denial-of-service.  \\ \hline
\end{tabularx}
\label{table:exp_effect}
\vspace{-4mm}
\end{table}

For each vulnerability, we decide its exploit effect based on the CVE description and confidentiality, integrity, and availability (CIA) subscores of CVSS. We first seek to extract the effect from the CVE description by matching the keywords for each effect type. If the description does not contain any keywords, we try to identify the \emph{exploit mechanism} defined in \cite{cwe} and communication protocol from the description using the same keyword matching approach. In combination with the CVSS' CIA subscores, we can infer the exploit effect. For example, suppose the exploit mechanism is buffer overflow. Then we check the CIA subscores to set the effect to \texttt{denial of service} (if only the availability score is greater than the threshold), or \texttt{root privilege} (if confidentiality, integrity, and availability are all greater than the threshold). 

The exploit models are also translated to Prolog facts. For example, the \texttt{vulProperty} fact in Listing \ref{list:example_exploit_model} is the exploit model for \texttt{CVE-2020-8864}. The precondition is the attacker being on the same Wifi network as \texttt{dLinkRouter}; the effect is that the attacker gets root privilege on this device. 

\begin{lstlisting}[caption={Prolog fact for an exploit model.},
label={list:example_exploit_model},]
vulExists(dLinkRouter, 'CVE-2020-8864').
vulProperty('CVE-2020-8864', wifiAdjacentLogically, rootPrivilege).
\end{lstlisting}

\subsection{Attack Graph Generator}
The Attack Graph Generator module takes the Prolog rule and fact file as input and verifies whether the attack goals (either provided by the administrator or automatically generated by \textsc{Iota}) can be achieved. If a goal can be achieved, it will generate the attack graph showing all the attack traces; otherwise, the IoT system is protected from that attack goal. When the administrator knows his security objectives, he can set the attack goal by taking the logic NOT of the objectives. For example, if the objective is to protect the camera from being rooted, then the attack goal is the attacker's gaining root privilege on the camera. When there are no security objectives specified, we enumerate all the potential privileges attackers may get on all the devices as attack goals. 

\begin{figure}[t]
\centering
\includegraphics[scale=0.117]{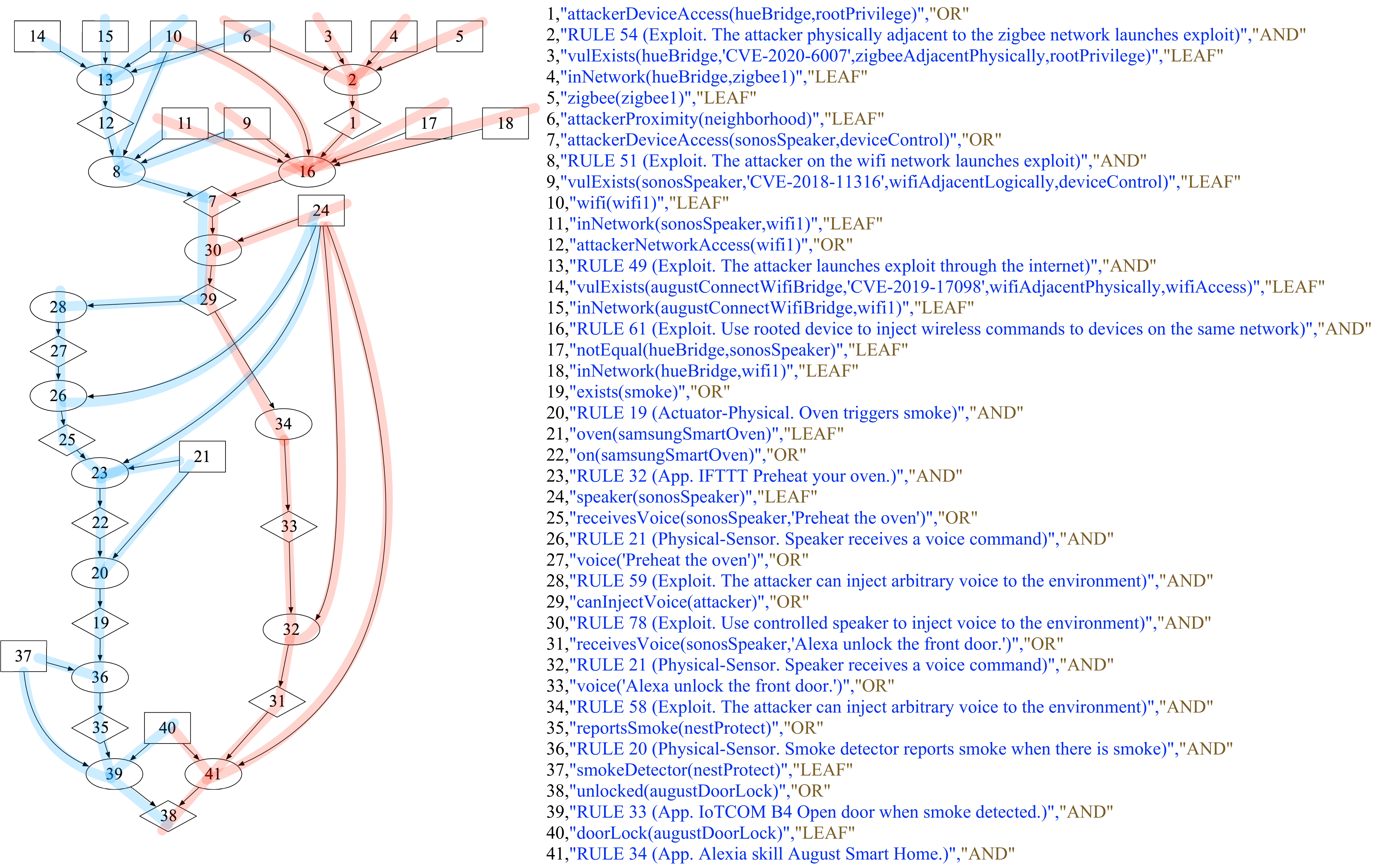}
\caption{IoT attack graph example. The meaning and type of each node is shown on the right.}
\label{fig:attack_graph_example}
\vspace{-2mm}
\end{figure}

Figure \ref{fig:attack_graph_example} shows a small attack graph, where node 38 represents the attack goal --- to unlock the doorlock. The meaning of each node is annotated on the right of the figure. In total, there are four \emph{attack traces} (formally defined in Section \ref{sec:analyzer}) in the attack graph, and two of them are highlighted in red and blue. The attacker can reach node 7 (i.e., controlling Sonos speaker) by exploiting Hue bridge or August Wifi bridge. And from node 7, there are two ways to get to node 38: via the Alexa skill \cite{alexa:august} (node 41) or by starting the oven to trigger smoke and using IoT app ``IoTCOM B4'' \cite{alhanahnah2020scalable} (node 39).

Essentially, there are three kinds of nodes. The rectangle nodes represent \emph{primitive facts} about the system state or the attacker state that are true before the exploit happens. The ellipse nodes represent Prolog \emph{rules}, such as exploits or apps' execution. The diamond nodes stand for \emph{derivation}, viz., new states about the system or the attacker after launching an exploit or executing an app. A derivation node can also be a precondition of another rule node. The logic meaning of each node is also annotated on the right of the figure. A detailed explanation of attack graph structure can be found in \cite{ou2006scalable}. 

%% file: sections/metrics.tex
\subsection{Attack Graph Analyzer}\label{sec:analyzer}

\setlength{\textfloatsep}{2pt}
\begin{algorithm}[t]
\SetAlgoLined\DontPrintSemicolon
\SetKwFunction{algo}{shortest\_trace}\SetKwFunction{proc}{shortest\_trace}
\SetKwRepeat{Do}{do}{while}%
\KwIn{(1) Attack graph $G$, (2) Attack goal node $n$}
\KwOut{Shortest attack trace to $n$ on $G$}
\SetKwProg{myalg}{Algorithm}{}{}
\myalg{\algo{$G, n$}}{
    $res\_node$ = TraceNode($n$)\;
    \If{$n$ \emph{is leaf node}}{
        \Return{\emph{(}$0, res\_node$\emph{)}}
    }
    \tcc{If current node is OR node, take the minimum of the parent nodes}
    \If{$n$ \emph{is OR node}}{
        Let $l$ be the list of parent nodes of $n$\;
        $min\_depth = \infty$\;
        \For{\emph{each node} $m$ \emph{in} $l$}{
            $(cur\_len, cur\_pred) = \proc{$G, m$}$\;
            \If{$min\_depth > $ cur\_len}{
                Set $res\_node.preds$ to $m$\;
                Update $min\_depth$\;
            }
        }
        \Return{$(min\_depth+1, res\_node)$}
    }
    \tcc{If current node is AND node, take the maximum of the parent nodes}
    \If{$n$ \emph{is AND node}}{
        Let $l$ be the list of parent nodes of $n$\;
        $min\_depth = -\infty$\;
        Set $res\_node.preds$ to $l$\;
        \For{\emph{each node} $m$ \emph{in} $l$}{
            $(cur\_len, cur\_pred) = \proc{$G, m$}$\;
            \If{$min\_depth < $ cur\_len}{
                Update $min\_depth$\;
            }
        }
        \Return{$(min\_depth+1, res\_node)$}
    }
}{}
\caption{Shortest Attack Trace Algorithm}
\label{alg:shortest}
\end{algorithm}

Because the generated attack graph can be gigantic, containing thousands of nodes, it is impractical to visualize the graph. Therefore, we propose two metrics to extract critical attack traces and quantify the impact of vulnerabilities.

\textbf{Shortest Attack Trace}. Among all of the attack traces to a specific attack goal, the \emph{shortest attack trace} takes the minimum number of exploits and provides a lower bound of the attack complexity to that goal node. For instance, the shortest attack trace to the goal node (node 5) in Figure \ref{fig:attack_graph_example} is highlighted in red whose depth is 12. Below we formally define the shortest attack trace and relevant concepts. 

\begin{definition*}{\textbf{(Attack Trace)}}
Given an attack graph $G$, an attack trace to a derivation node $n$ is a subgraph $G^\prime$ satisfying the following conditions: (1) Any OR node of $G^\prime$ has only one incoming edge; (2) Any AND nodes of $G^\prime$ has incoming edges from all its parent nodes; (3) All the source nodes of $G^\prime$ are primitive fact nodes; and (4) The sink node of $G^\prime$ is node $n$.
\end{definition*}

\begin{definition*}{\textbf{(Depth of an Attack Trace)}}
The depth of an attack trace is the \emph{longest path} from any primitive fact node to the sink node of the attack trace. 
\end{definition*}

\begin{definition*}{\textbf{(Shortest Attack Trace)}}
For a given attack graph and a derivation node $n$, the shortest attack trace is the attack trace to $n$ with the smallest depth.
\end{definition*}

We cannot apply Dijkstra's algorithm to the shortest attack trace problem because our definition of shortest attack trace is different from the shortest path in graph theory: (1) There can be multiple source nodes; (2) The attack trace is a subgraph, not a path. Hence, we design a recursive algorithm, i.e., Algorithm \ref{alg:shortest}, to compute the shortest attack trace to a specified attack goal node. The depth of a leaf node is defined as 0.

\begin{figure}[t]
\centering
\includegraphics[scale=0.3]{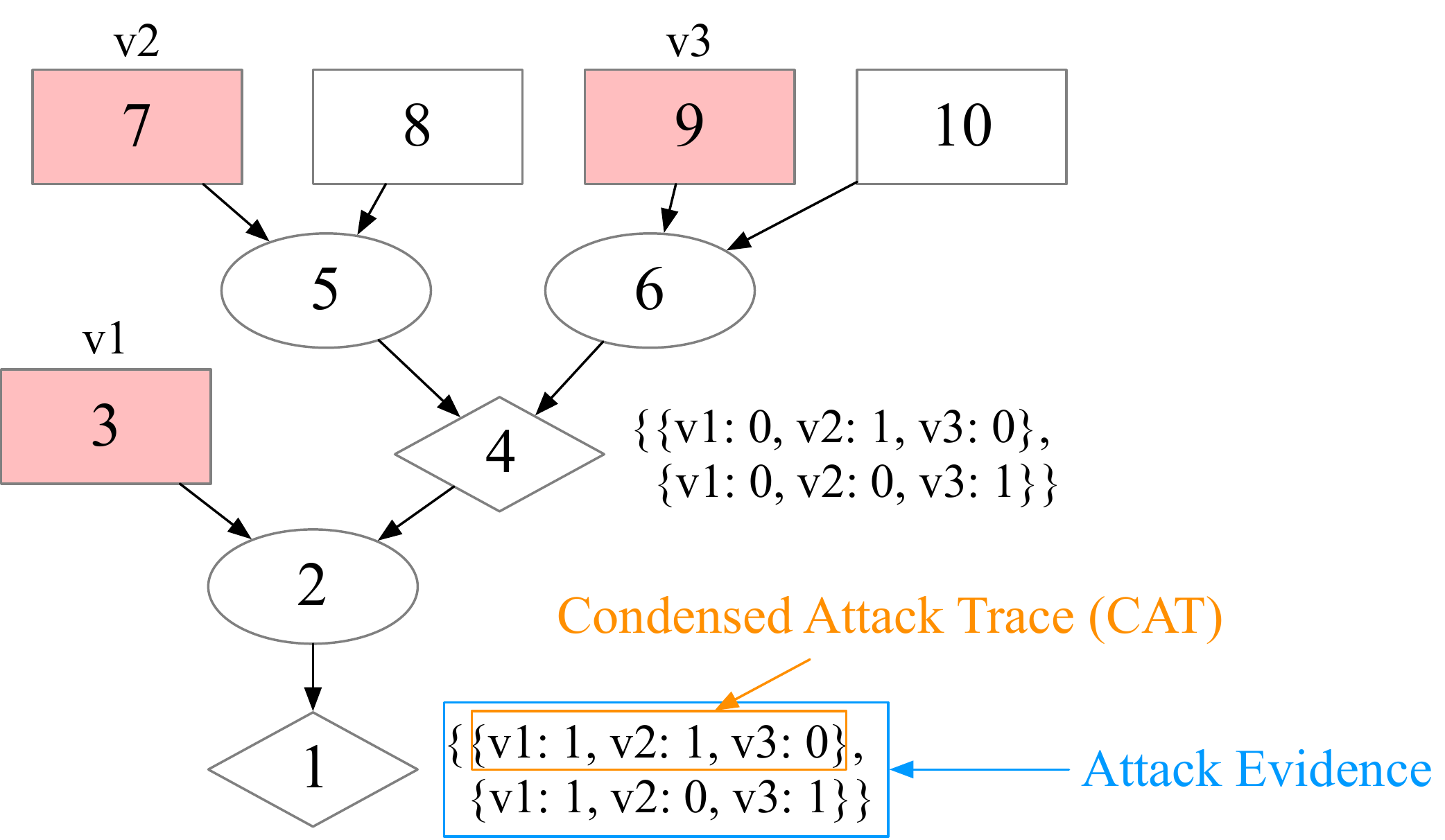}
\caption{Example attack graph and the corresponding attack evidence for node 1 and node 4. Node 3, 7, and 9 are primitive fact nodes describing different vulnerabilities represented as $v1$, $v2$, and $v3$.}
\label{fig:vul_evi_example}
\end{figure}

\textbf{Blast Radius}. The \emph{blast radius} measures each vulnerability's impacts on the IoT system and can be used for system hardening. For example, if vulnerability A's blast radius is a superset of that of vulnerability B, we conclude that A's impact is bigger than B's, and therefore we should fix A first. 

\begin{definition*}{\textbf{(Blast Radius (BR))}}
Given an attack graph, the blast radius of vulnerability $v$ is the set of all of the privileges (represented as derivation nodes) the attacker gets after exploiting \emph{only} $v$.
\end{definition*}

As there can be more than one trace to a certain node, and a vulnerability can be used in multiple attacks, we must keep track of vulnerabilities involved for each trace to a certain node in the attack graph. We come up with the following concepts to help us compute the blast radius of each vulnerability.

\begin{definition*}{\textbf{(Condensed Attack Trace (CAT))}}
Given an attack graph $G$, the condensed attack trace of a node $n$ is the map from all of the vulnerabilities on $G$ to 0 (when the vulnerability is not used) or 1 (when used) along some attack trace to $n$.
\end{definition*}

\begin{definition*}{\textbf{(Attack Evidence)}}
The attack evidence of a node $n$ is the set of its condensed attack traces.
\end{definition*}

\begin{algorithm}[h]
\SetAlgoLined\DontPrintSemicolon
\SetKwFunction{algo}{merge\_ae\_OR}\SetKwFunction{proc}{merge\_ae}
\SetKwRepeat{Do}{do}{while}
\KwIn{Attack Evidence of two nodes: $a$, $b$}
\KwOut{Attack Evidence of the child node $c$, an OR node}
\SetKwProg{myalg}{Algorithm}{}{}
\myalg{\algo{$a, b$}}{
    Let $merged\_ae$ be a copy of $a.ae$\;
    \For{$cat$ \emph{in} $b.attack\_evidence$}{
        \If{\emph{$cat$ not in $a.attack\_evidence$}}{
            $merged\_ae$.append($cat$)
        }
    }
    \Return{$merged\_ae$}\;
}{}
\caption{Attack Evidence Merge --- OR}
\label{alg:vul_ev_merge_OR}
\end{algorithm} 


\begin{algorithm}[h]
\SetAlgoLined\DontPrintSemicolon
\SetKwFunction{algo}{merge\_ae\_AND}\SetKwFunction{proc}{merge\_ae}
\SetKwRepeat{Do}{do}{while}
\KwIn{(1) Attack evidence of two nodes: $a$, $b$, (2) $Vuls$: The set of all the CVEs on the attack graph}
\KwOut{Attack evidence of the child node $c$, an AND node}
\SetKwProg{myalg}{Algorithm}{}{}
\myalg{\algo{$a, b, Vuls$}}{
    $merged\_ae$ = []\;
    \For{$cat1$ \emph{in} $a.attack\_evidence$}{
        \For{$cat2$ \emph{in} $b.attack\_evidence$}{
            \tcc{Initialize $merged$, suppose $\abs{Vuls} = p$}
            $merged$ = $\{v_1: 0 ,\dots, v_p: 0\}$\;
            \lFor{\emph{$vul$ in $Vuls$}}{
                $merged$[$vul$] = max($cat1$[$vul$], $cat2$[$vul$])
            }
            \If{\emph{$merged$ not in $merged\_ae$}}{
                $merged\_ae$.append($merged$)
            }
        }
    }
    \Return{$merged\_ae$}\;
}{}
\caption{Attack Evidence Merge --- AND}
\label{alg:vul_ev_merge_AND}
\end{algorithm} 

Figure \ref{fig:vul_evi_example} illustrates an example attack graph and the corresponding attack evidences for node 1 and node 4. Since there are two attack traces to node 4 involving different vulnerabilities, the attack evidence for node 4 contains two elements, so is node 1. We compute the vulnerability evidence for each node in a forward fashion from leaf nodes to the goal nodes. Our merging algorithms are explained in Algorithm \ref{alg:vul_ev_merge_OR} and Algorithm \ref{alg:vul_ev_merge_AND} for OR and AND nodes, respectively. After getting the vulnerability evidence for each node, we can determine whether a derivation node should be included in some vulnerability's blast radius using Algorithm \ref{alg:determine_blast_radius}. The complete blast radius algorithm is given in Algorithm \ref{alg:blast}.

\begin{algorithm}[t]
\SetAlgoLined\DontPrintSemicolon
\SetKwFunction{algo}{determine\_br}\SetKwFunction{proc}{determine\_br}
\SetKwRepeat{Do}{do}{while}%
\KwIn{(1) $att\_ev$: attack evidences for all of the attack graph nodes, (2) $Vuls$: the map from $node\_id$ to the node's vulnerability evidence for all of the nodes}
\KwOut{The BR of each vulnerability in the attack graph}
\SetKwProg{myalg}{Algorithm}{}{}
\myalg{\algo{$att\_ev$, $Vuls$}}{
    \tcc{initialize $br$}
    Let $br$ be an empty map\;
    \For{\emph{$vul$ in $Vuls$}}{
        $br[vul]$ = $\emptyset$
    }
    \For{$n$ \emph{in} $att\_ev$}{
        \If{\emph{$n$.type is OR}}{
            \For{$cat$ \emph{in} $att\_ev[n]$}{
                \If{\emph{sum($cat$.values()) == 1}}{
                    Find the $key$ s.t. $cat$[$key$] == 1\;
                    $br$[$key$] = $br$[$key$] $\cup\enspace \{n\}$
                }
            }
        }
    }
    \Return{\emph{$br$}}\;
}{}
\caption{Determine Blast Radius}
\label{alg:determine_blast_radius}
\end{algorithm}

\begin{algorithm}[t]
\SetAlgoLined\DontPrintSemicolon
\SetKwFunction{algo}{blast\_radius}
\SetKwRepeat{Do}{do}{while}
\KwIn{Attack graph $G$}
\KwOut{Blast radius of each vulnerability in $G$}
\SetKwProg{myalg}{Algorithm}{}{}
\myalg{\algo{$G$}}{
    \tcc{Generate the list of unique vulnerabilities}
    $Vuls$ = []\;
    \For{\emph{$node$ in $G$}}{
        \If{\emph{$node$ is a primitive fact node \textbf{and} $node$ describes a vulnerability $v$}}{
            $Vuls$.append($v$)\;
        }
    }
    \tcc{Initialize attack evidence for primitive fact nodes, suppose $\abs{Vuls} = p$}
    $queue$ = []\;
    \For{\emph{$node$ in $G$}}{
        $node$.$cat$ = [$\{v_1: 0, \dots, v_p: 0\}$]\;
        \If{\emph{$node$ is a primitive fact node \textbf{and} $node$ describes a vulnerability $v$}}{
            find the index $i$ such that $Vuls[i]$ = $v$\;
            $node$.$cat$ = [$\{v_1: 0, \dots, v_i: 1, \dots, v_p: 0\}$]\;
            \For{\emph{$child$ in $node$.children}}{
                \If{\emph{$child$ not in $queue$}}{
                    $queue$.enqueue($child$)\;
                }
            }
        }
    }
    \tcc{Iteratively build attack evidence for all the nodes}
    \While{\emph{$queue$.length != 0}}{
        $node$ = $queue$.dequeue()\;
        $cur\_ae$ = $node$.parents[0].cat\;
        \For{\emph{$i$ in 1 to length($node$.parents)}}{
            $next\_ae$ = $node$.parents[i].cat\;
            \If{\emph{$node$.type is AND}}{
                $cur\_ae$ = \texttt{merge\_ae\_AND}($cur\_ae$, $next\_ae$)\;
            }
            \Else{
                $cur\_ae$ = \texttt{merge\_ae\_OR}($cur\_ae$, $next\_ae$)\;
            }
        }
        $node$.cat = $cur\_ae$\;
    }
    Let $att\_ev$ be the attack evidences for all nodes\;
    \Return{\emph{\texttt{determine\_br($att\_ev$, $Vuls$)}}}\;
}{}
\caption{Blast Radius Algorithm}
\label{alg:blast}
\end{algorithm}

Attack evidence provides a summary of vulnerabilities involved along each attack trace to a certain node and is useful for other important problems. For example, we can use it to compute the \emph{minimal set of vulnerabilities to patch to thwart an attack goal} defined in \cite{sheyner2002automated}. We can count the occurrence of each vulnerability in the attack evidence and iteratively choose the vulnerabilities in descent order of occurrence; if the current vulnerability is in the same condensed attack trace of some chosen vulnerability, then we consider the next one. The iteration stops when all of the condensed attack traces contain at least one vulnerability chosen. For another application, if the administrators have assigned numerical values as the \emph{complexity} of each exploit, they can calculate the complexity of each attack trace by summing the exploit complexity for each condensed attack trace.

%% file: sections/implementation.tex
\section{Implementation} \label{sec:implementation}
The \textsc{Iota} modules are implemented in Python using 2475 LoC. Physical dependencies and exploit rules are implemented in Prolog using 1179 LoC. The framework utilizes MySQL Connector Python library \footnote{https://github.com/mysql/mysql-connector-python} for database operations and MulVAL \cite{ou2005mulval} for attack graph generation. 

\textbf{Translator}. The Translator module converts IoT system configuration and vulnerabilities to Prolog clauses. The initial \textbf{system configuration} (specified in JSON format) is sent to the Translator module to generate Prolog facts. The example system configuration and the translated results are shown in Listing \ref{list:sys_input} and \ref{list:sys_prolog}, respectively. 

\begin{lstlisting}[
    firstnumber=1,
    caption={Example IoT system configuration JSON file.},
    label={list:sys_input}]
{
    "devices": [
        {"name": "D-Link Router",
         "type": "router",
         "network": ["wifi1"]
        },
        {"name": "Smartthings Hub",
         "type": "gateway",
         "network": ["wifi1", "zigbee1"]
        }
    ],
    "networks": [
        {"name": "wifi1",
         "type": "Wifi"
        },
        {"name": "zigbee1",
         "type": "Zigbee"
        }
    ]
}
\end{lstlisting}
The above JSON file lists the device and network settings of an IoT system. The device and network names are specified by the user, while device and network types use standard names predefined. 

\begin{lstlisting}[
    caption={Translated Prolog facts on system configuration.},
    label={list:sys_prolog}]
router(dLinkRouter).
inNetwork(dLinkRouter, wifi1).

gateway(smartthingsHub).
inNetwork(smartthingsHub, wifi1).
inNetwork(smartthingsHub, zigbee1).

wifi(wifi1).
zigbee(zigbee1).
\end{lstlisting}

\textbf{IoT apps} are first sent to the App Semantic Extractor and then translated to Prolog rules. Listing \ref{list:app_config} is an example configuration of the SmartApp \textit{Hall Light} explained in Section \ref{sec:app_logic_extractor}. The Translator combines parsed app semantic tuple (Listing \ref{list:app_logic_tuple}) and app configuration (Listing \ref{list:app_config}) to generate Prolog rules shown in Listing \ref{list:example_iot_app}.

\begin{lstlisting}[
    basicstyle=\ttfamily\footnotesize,
    caption={Example IoT app configuration JSON file.},
    label={list:app_config}]
{
  "apps": [
    {"App name": "Light on when I come home",
     "description": "Turn on the hall light if           there is motion and the door opens.",
     "device map": {
        "bulb": "Hue Wifi Bulb",
        "contact sensor": "Ring Contact Sensor",
        "motion sensor": "Mijia Motion Sensor"
      }
    }
  ]
}
\end{lstlisting}

\begin{lstlisting}[style=base,
caption={Prolog rule for IoT app \textit{Hall Light: Welcome Home}.},
label={list:example_iot_app},]
on(Bulb) :- 
    bulb(Bulb),
    reportsMotion(MotionSensor),
    motionSensor(MotionSensor),
    open(DoorContactSensor),
    doorContactSensor(DoorContactSensor).
\end{lstlisting}

\textbf{Attack Graph Generator}.
The Attack Graph Generator concatenates exploit rules, indirect physical dependency rules, and the translated IoT app rules into a Prolog rule file. It also combines all the translated Prolog facts (including facts about device and network configuration and direct physical dependencies) and vulnerabilities (i.e., vulnerability existence facts and exploit model facts) into a Prolog fact file. The attack goals are also inserted into the Prolog fact file. After that, the rule and fact file are then sent to MulVAL \cite{ou2005mulval} library to generate the Prolog reasoning log file and the attack graph. 

%% file: sections/evaluation.tex
\section{Evaluation}\label{sec:eval}

\subsection{Dataset}
To generate attack graphs and conduct analysis, we need to obtain IoT system configurations, including device instances and IoT apps installed. Though thousands of IoT apps are available, how users choose apps and device instances to install is still unknown. To the best of our knowledge, currently, there is no public dataset of IoT systems configured by different users. Such information gap has long been a challenge to IoT system security research \cite{nguyen2018iotsan, alhanahnah2020scalable}. To evaluate \textsc{Iota}, we generate synthetic IoT systems based on real-world IoT apps and device instances. We use a top-down approach to generate IoT systems by choosing the IoT app bundles first, as they determine the whole system's functionality. Once we have determined the IoT app bundle for the system, we create system instances by selecting device instances. To emulate the scenario where a user installs IoT devices but does not connect them to any IoT apps, we add individual IoT devices in one-third of the system instances created.

We consider SmartApps in the SmartThings Repository \footnote{https://github.com/SmartThingsCommunity/SmartThingsPublic}, and IFTTT applets \footnote{https://ifttt.com/search/query/smart\%20home?tab=applets} for SmartThings platform and create a pool of 532 IoT apps. We build a list of 59 smart home devices of 26 types, covering all of the device types listed on SmartThings Products List \footnote{https://www.smartthings.com/products-list}, from motion sensors, outlets to home appliances like TV, smart oven, etc. The devices are from 16 different platforms, all of which, except Roku, HP, and Aqara, are listed on Smartthing Partners\footnote{https://www.smartthings.com/partners}. In total, we create 37 IoT system instances. The first 18 instances are created based on the 6 app bundles used in \cite{alhanahnah2020scalable} (which contains malicious apps), while the next 12 instances are generated based on 4 app bundles chosen from our app pool (which are treated as benign apps). The last 7 systems are of bigger size, with at most 50 devices, to further evaluate the scalability of our framework. 

\begin{figure}[t]
\centering
\includegraphics[scale=0.35]{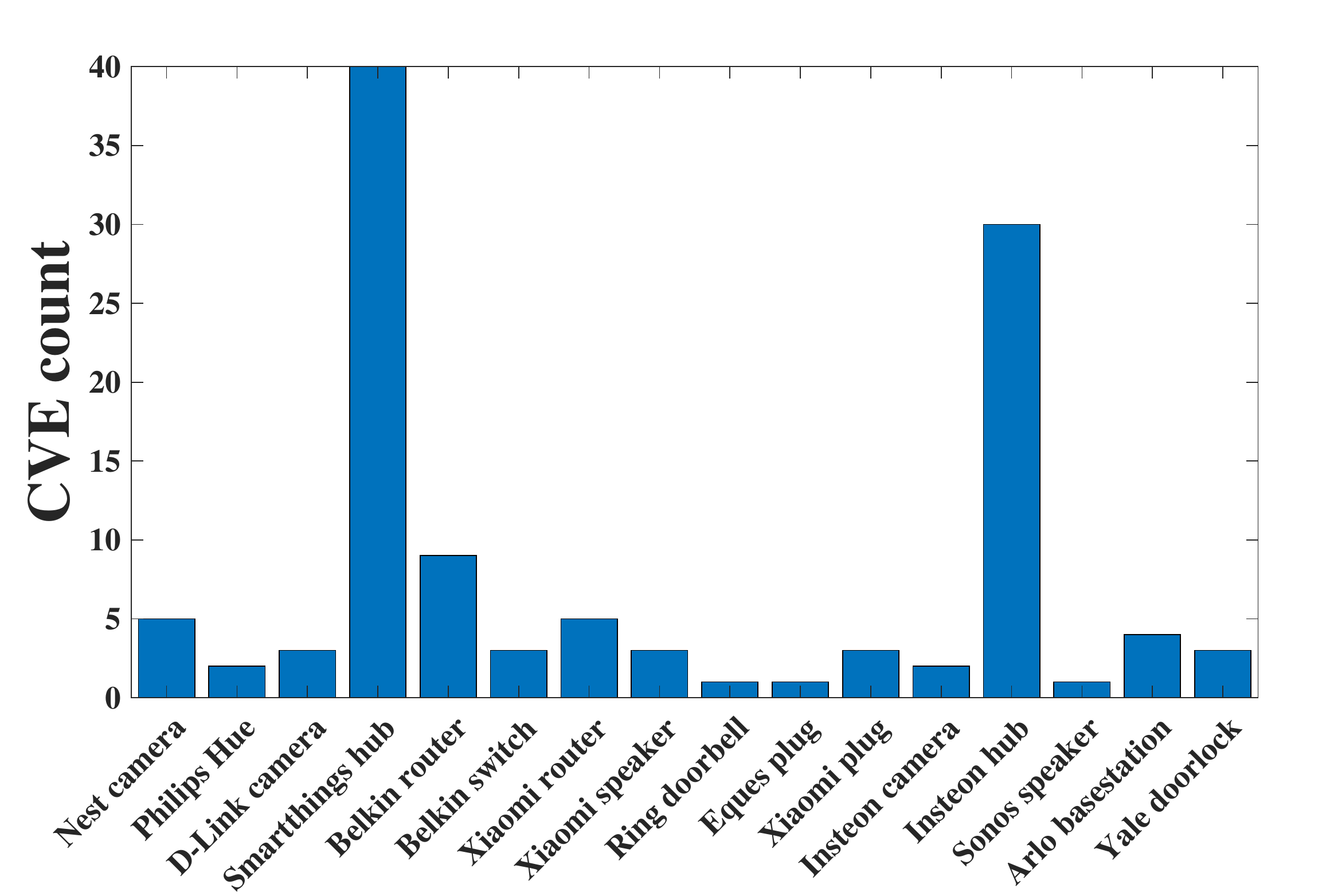}
\caption{The number of CVEs scanned on IoT devices.}
\label{fig:query}
\end{figure}

\subsection{Results}
\textbf{Vulnerability Scanning}. 
The vulnerability scanner queries CVE database with the full name of a given IoT device instance. The scanning result is shown in Figure \ref{fig:query}, where the devices with the largest number of CVEs are illustrated. On average, there are 7.2 CVEs per device. Figure \ref{fig:query} shows that device types with the largest number of detected vulnerabilities are routers, cameras, and gateways. The reason could be that due to their pivotal position in IoT systems, security researchers tend to analyze these types of devices. We manually checked all CVE records and found out that $94.2\%$ of them are relevant to the queried device. The typical devices and their CVEs identified are listed in Table \ref{table:device_cve}.

We further verified the scanned CVEs with 12 real-world IoT devices and found out 5 of them still contain vulnerabilities: we obtained the exploit scripts for Philips Wifi Bulb, D-Link DCS-5009L Camera, and Eques Elf Smart Plug and successfully launched attacks against these devices. For Wemo Insight Smart Plug and Radio Thermostat, we confirmed the existence of the vulnerabilities by matching the firmware version of devices with the one in the vulnerability reports.

\begin{table}[t]
\small
\centering
\caption{CVEs on typical IoT devices.}
\begin{tabular}{|M{3cm}|M{4.9cm}|}
\hline
\rowcolor[HTML]{C0C0C0}
\textbf{Device Instance} & \textbf{Typical CVE(s) Scanned} \\\hline
Hue Wifi Bulb & CVE-2019-18980 \\\hline
Hue Bridge & CVE-2020-6007 \\\hline
Nest Cam IQ Indoor & CVE-2019-5035, CVE-2019-5036, CVE-2019-5037 \\\hline
D-Link DCS Camera & CVE-2019-10999 \\\hline
Ring Doorbell & CVE-2019-9483 \\\hline
Yale Lock & CVE-2019-17627 \\\hline
August Bridge & CVE-2019-17098 \\\hline
Smartthings Hub & CVE-2018-3904, CVE-2018-3917, 	CVE-2018-3919, CVE-2018-3925 \\\hline
Xiaomi Gateway & CVE-2019-15913, CVE-2019-15914 \\\hline
Hue Bridge & CVE-2020-6007 \\\hline
Arlo Basestation & CVE-2019-3949, CVE-2019-3950 \\\hline
Sonos Speaker & CVE-2018-11316 \\\hline
Xiaomi Motion Sensor & CVE-2019-15913 \\\hline
\end{tabular}
\label{table:device_cve}
\vspace{-2mm}
\end{table}

\begin{figure}[t]
\centering
    \begin{subfigure}[t]{0.49\columnwidth}
        \includegraphics[width=\linewidth]{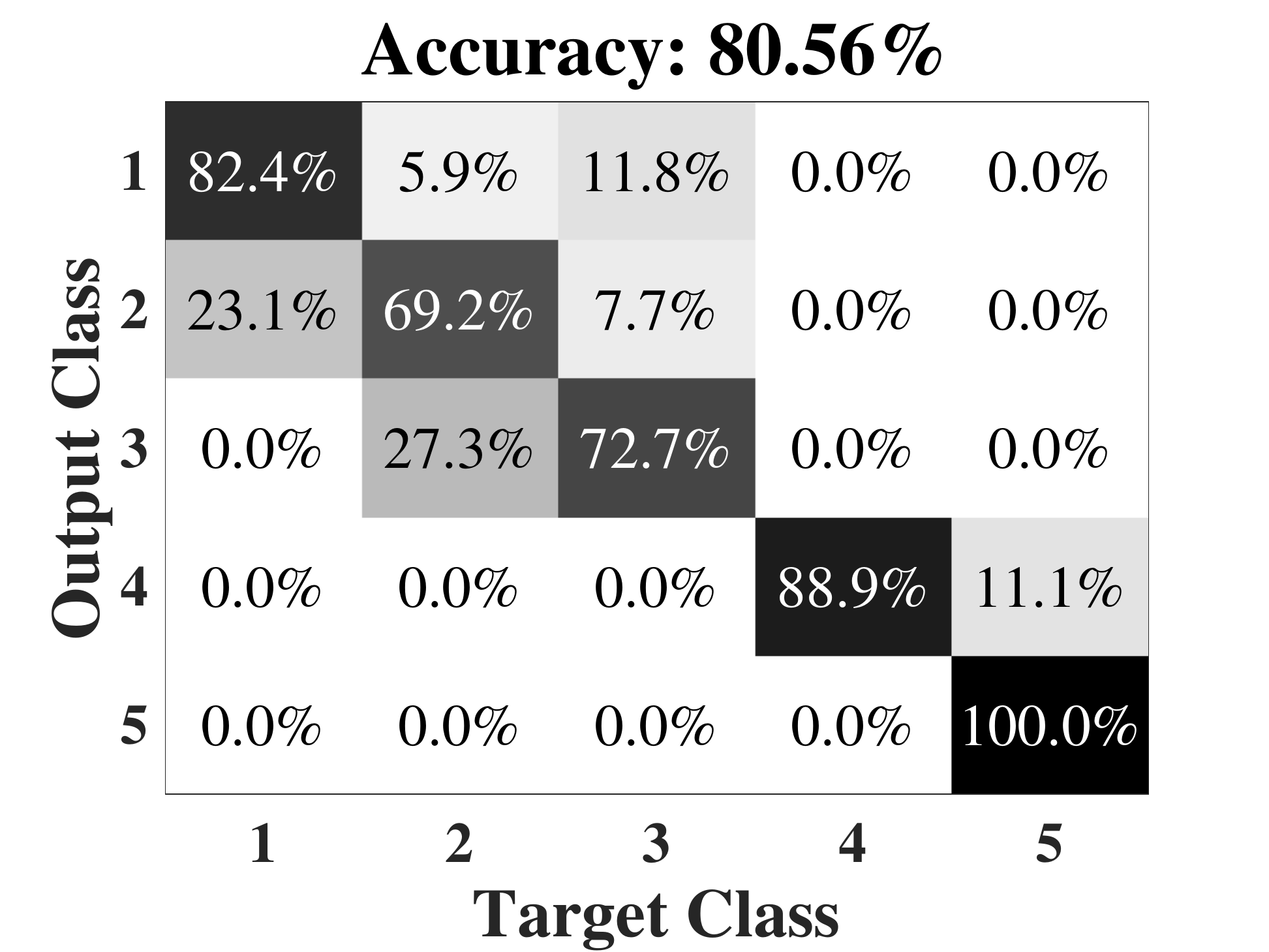}
        \caption{}
        \label{fig:precondition}
    \end{subfigure}
    \begin{subfigure}[t]{0.49\columnwidth}
        \includegraphics[width=\linewidth]{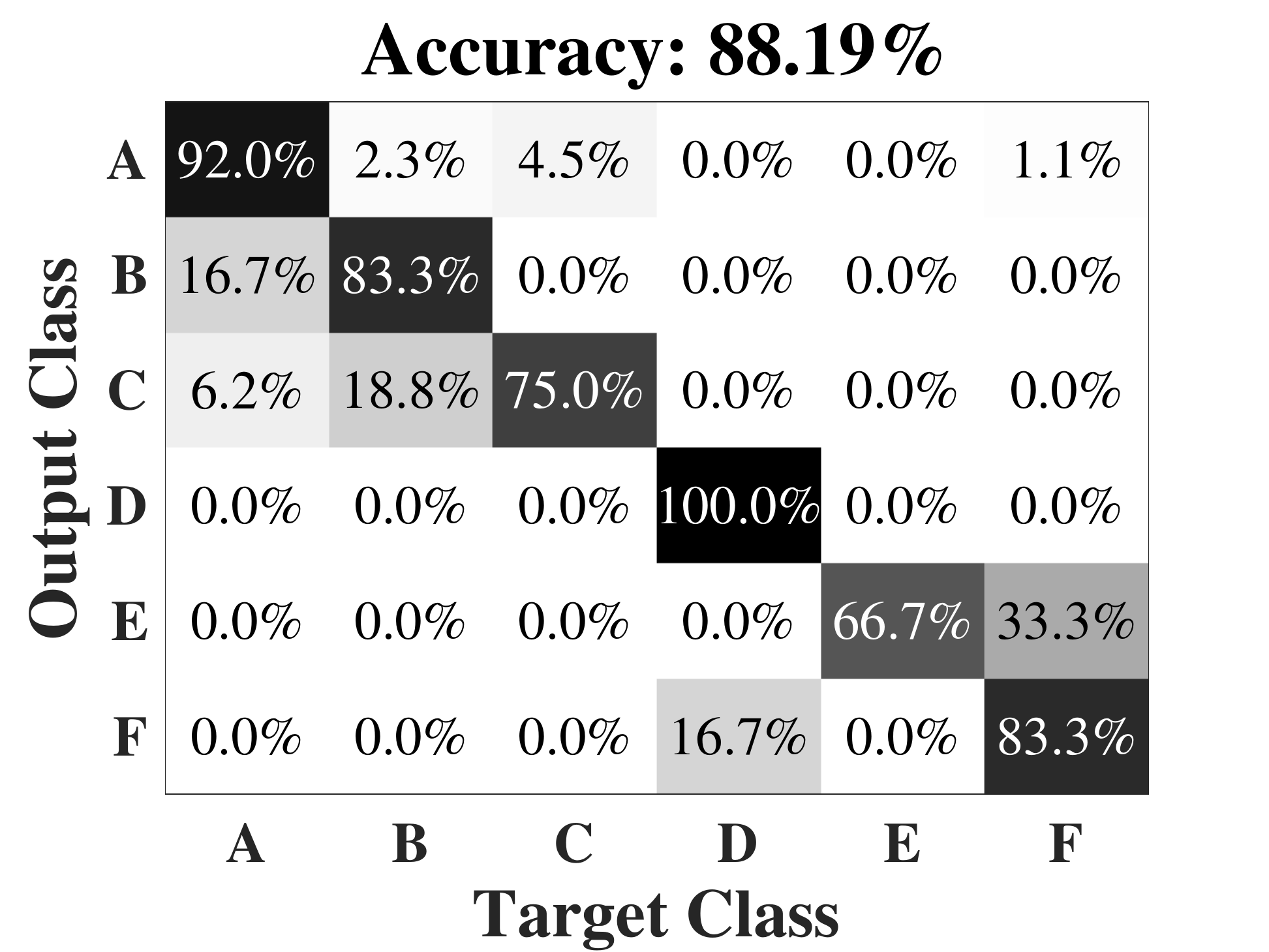}
        \caption{}
        \label{fig:effect}
    \end{subfigure}
    \caption{(a) Confusion matrix for exploit precondition identification. Label \textbf{1} to \textbf{5} denote preconditions listed in Table \ref{table:exp_precondition}. (b) Confusion matrix for exploit effects. Label \textbf{A} to \textbf{F} represent exploit effects listed in Table \ref{table:exp_effect}.}
    \label{fig:confusion}
\end{figure}

\textbf{Vulnerability Analysis}. We ran our vulnerability analyzer on 127 CVE records of smart home IoT devices collected by the Vulnerability Scanner module and manually checked the accuracy of the predicted exploit precondition and effects. The results are shown in Figure \ref{fig:confusion}. Overall, our Vulnerability Analyzer achieves 80\% and 88\% prediction accuracy for precondition and effect, respectively. From Figure \ref{fig:confusion}(a), we can see that the class \texttt{local} and \texttt{physical} have the highest accuracy, because the CVSS attack vectors for \texttt{physical} type is almost 100\% accurate. And for low-power protocols, most of the time, the exploit range is local; hence, we can decide the \texttt{local} type with the help of protocol type. The precondition types with the lowest prediction accuracy are \texttt{Adjacent physically} and \texttt{Adjacent logically}. This is because some CVEs' descriptions provide vague information for these two types. 

\begin{table*}[t]
\small
\begin{threeparttable}
\centering
\caption{Attack graph analysis results on IoT system instances.}
\begin{tabular}{|M{1.5cm}|M{1.4cm}|M{1.4cm}|M{1.4cm}|M{1.4cm}|M{1.4cm}|M{2.5cm}|M{3.6cm}|}
\hline
\rowcolor[HTML]{C0C0C0}
\textbf{System ID} & \textbf{\# Devices} & \textbf{\# CVEs} & \textbf{\# Nodes} & \textbf{\# Edges} & \textbf{\# Goals} & \textbf{Shortest depth$^{*}$} & \textbf{CVE ($\abs{\text{BR}}$)$^{\dagger}$} \\\hline
1 & 4 & 3 & 12 & 15 & 1 & 6 & CVE-2019-18980 (4) \\\hline
4 & 6 & 4 & 37 & 39 & 5 & (2, 8) & CVE-2020-6007 (5) \\\hline
8 & 7 & 4 & 35 & 44 & 4 & (4, 10) & CVE-2019-18980 (4) \\\hline
11 & 7 & 3 & 25 & 26 & 9 & (6, 16) & CVE-2018-3904 (9) \\\hline
19 & 10 & 6 & 36 & 35 & 4 & (2, 8) & CVE-2020-6007 (4) \\\hline
26 & 12 & 7 & 117 & 173 & 19 & (4, 10) & CVE-2020-6007 (5) \\\hline
28 & 15 & 9 & 130 & 182 & 20 & (4, 18) & CVE-2019-3949 (29) \\\hline
32 & 23 & 11 & 131 & 186 & 23 & (2, 8) & CVE-2018-3904 (12) \\\hline 
33 & 31 & 16 & 209 & 310 & 23 & (2, 10) & CVE-2018-3904 (19) \\\hline 
37 & 50 & 28 & 338 & 577 & 43 & (2, 14) & CVE-2018-11314 (32) \\\hline
\end{tabular}
\begin{tablenotes}
    \small
    \item $^{*}$: When there are multiple attack goals in attack graph, $(x, y)$ means the min and max depth of the shortest attack traces to different goals.
    \item $^{\dagger}$: $\abs{\text{BR}}$ is the cardinality of blast radius of the CVE. This column shows the CVE with the largest blast radius cardinality.
\end{tablenotes}
\label{table:eval_graph_analysis}
\end{threeparttable}
\end{table*}

According to Figure \ref{fig:confusion}(b), the most accurate class is \texttt{root}. This is because there are multiple effective indicators, such as keywords like ``root'', ``arbitrary'', etc., the CVSS subscores (confidentiality, integrity, and availability subscore all being high), and the exploit mechanism like buffer overflow, or integer overflow, etc. With these indicators combined, our prediction is accurate. The high accuracy for both the precondition and effect prediction shows our module is highly effective. 

\textbf{Attack Graph Generation and Analysis}. Table \ref{table:eval_graph_analysis} illustrates analysis results for 10 IoT system instances from the 37 instances we built. The first column is the ID of the system. The first four rows are the analysis results for systems built based on app bundles used in \cite{alhanahnah2020scalable}, and the rest of the rows are results for systems built from our own app bundles. The \textbf{\# CVEs} column is the number of vulnerabilities found on the given system. We enumerate all of the system resource compromises as potential attack goals, and the \textbf{\# Goals} column denotes the number of attack goals shown on the attack graph, which can be achieved by the attacker for a given system. 

Table \ref{table:distribution} shows the distribution of the shortest depths of the attack traces to different goal nodes for 10 attack graphs in Table \ref{table:eval_graph_analysis}. From the figure, the largest portion (43.9\%) of the attack traces have the shortest depths among 5 $\sim$ 8. To evaluate the effectiveness of the attack graphs, we manually check 27 shortest attack traces whose depths are at least 9. As a result, $62.8\%$ of the attack traces revealed by \textsc{Iota} are not anticipated by the system designers.

\begin{table}[t]
\small
\centering
\caption{Distribution of the shortest depths for different attack goals. $\bm{d}$ means the depth of the shortest attack trace to an attack goal node.}
\begin{tabular}{|M{3cm}|M{2cm}|M{2cm}|}
\hline
\rowcolor[HTML]{C0C0C0}
\textbf{Shortest Trace Depth} & \textbf{Trace Count} & \textbf{Percentage}\\\hline
$d\leq 4$ & 51 & 36.7\% \\\hline
$5\leq d \leq 8$ & 61 & 43.9\% \\\hline
$9\leq d \leq 12$ & 24 & 17.3\% \\\hline
$13\leq d \leq 16$ & 2 & 1.4\% \\\hline
$d \geq 17$ & 1 & 0.7\% \\\hline
\end{tabular}
\label{table:distribution}
\vspace{-2mm}
\end{table}

\textbf{Case study}. \emph{System 37} in Table \ref{table:eval_graph_analysis} consists of 50 different devices, including all of the device types in Figure \ref{fig:query}, and Wifi printer, smart TV, humidifier and toaster, etc. The vulnerability \texttt{CVE-2018-11314} identified on the Roku TV has the largest blast radius, whose cardinality is 32. By exploiting \texttt{CVE-2018-11314}, the attacker on the internet can control the smart TV and play arbitrary video. System 37 has multiple voice-related IoT apps installed, such as turning on/off the light, turning on/off the humidifier, opening the window, and locking/unlocking the door if the smart speaker receives the corresponding voice command. As a result, after compromising the TV, the attacker can control those end devices by playing videos containing the voice commands. The attacker can further compromise physical environment features such as illuminance and humidity. The blast radius of \texttt{CVE-2018-11314} directly tells system administrators about all these compromises caused by this vulnerability.


As another example, we describe the shortest attack trace to the attack goal node ``opening the window'' in System 28, whose depth is 18. In this example, a physically adjacent attacker first exploits \texttt{CVE-2019-17098} on the smart lock gateway to sniff the home Wifi credentials. Then he exploits \texttt{CVE-2019-3949} on the camera basestation to control the indoor camera. After that, he utilizes the rooted camera to inject the voice command ``preheat the oven'' into the smart home, which is sensed by the smart speaker. The speaker triggers the IoT app to start the oven. The oven may trigger smoke, which is sensed by a smoke detector. Finally, another IoT app opens the window when smoke is detected. 

\subsection{Scalability}
\begin{figure}[t]
    \centering
    \begin{subfigure}[t]{0.235\textwidth}
        \centering
        \includegraphics[scale=0.21]{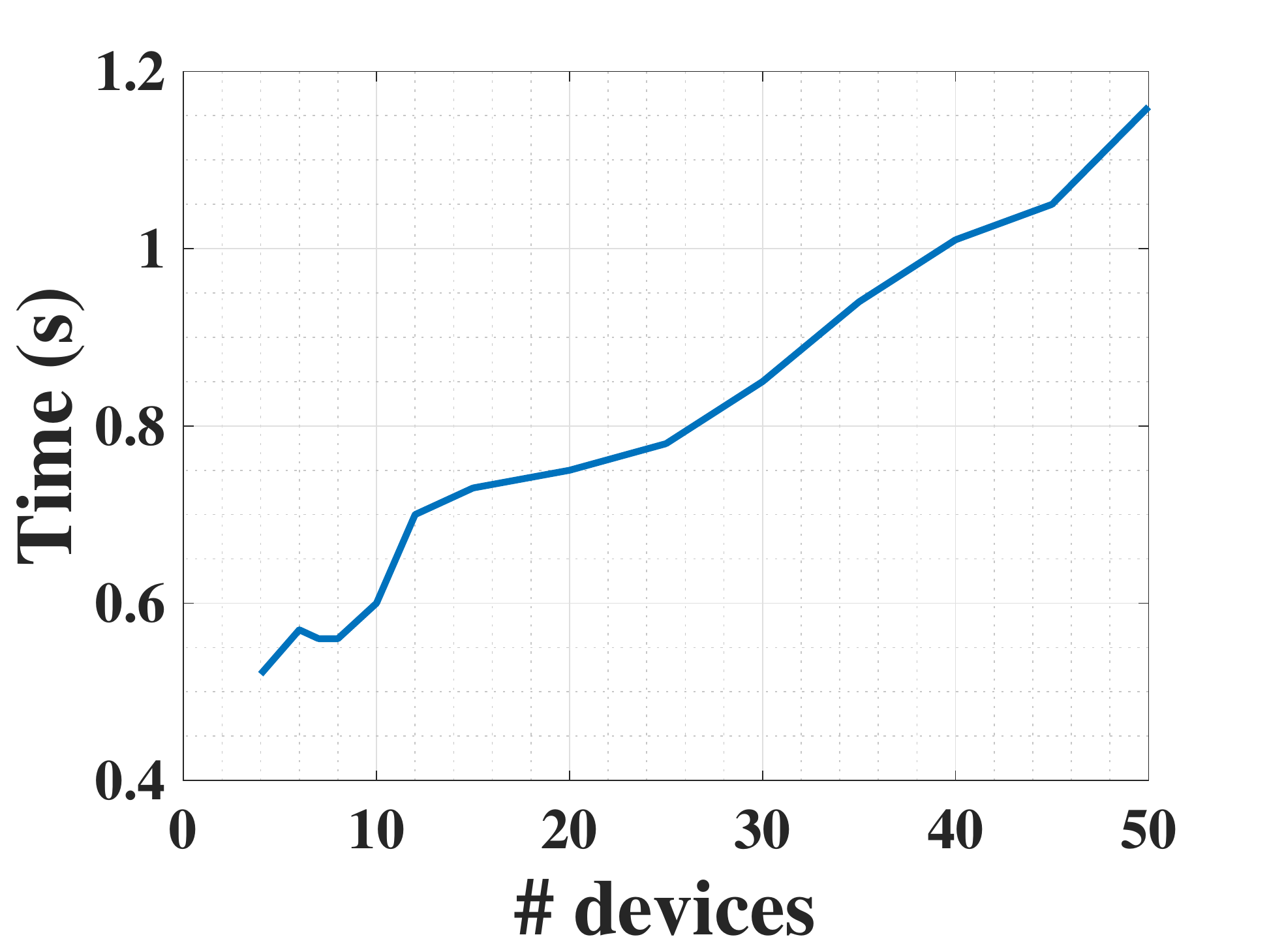}
        \caption{}
        \label{fig:time}
    \end{subfigure}
    \begin{subfigure}[t]{0.235\textwidth}
        \centering
        \includegraphics[scale=0.21]{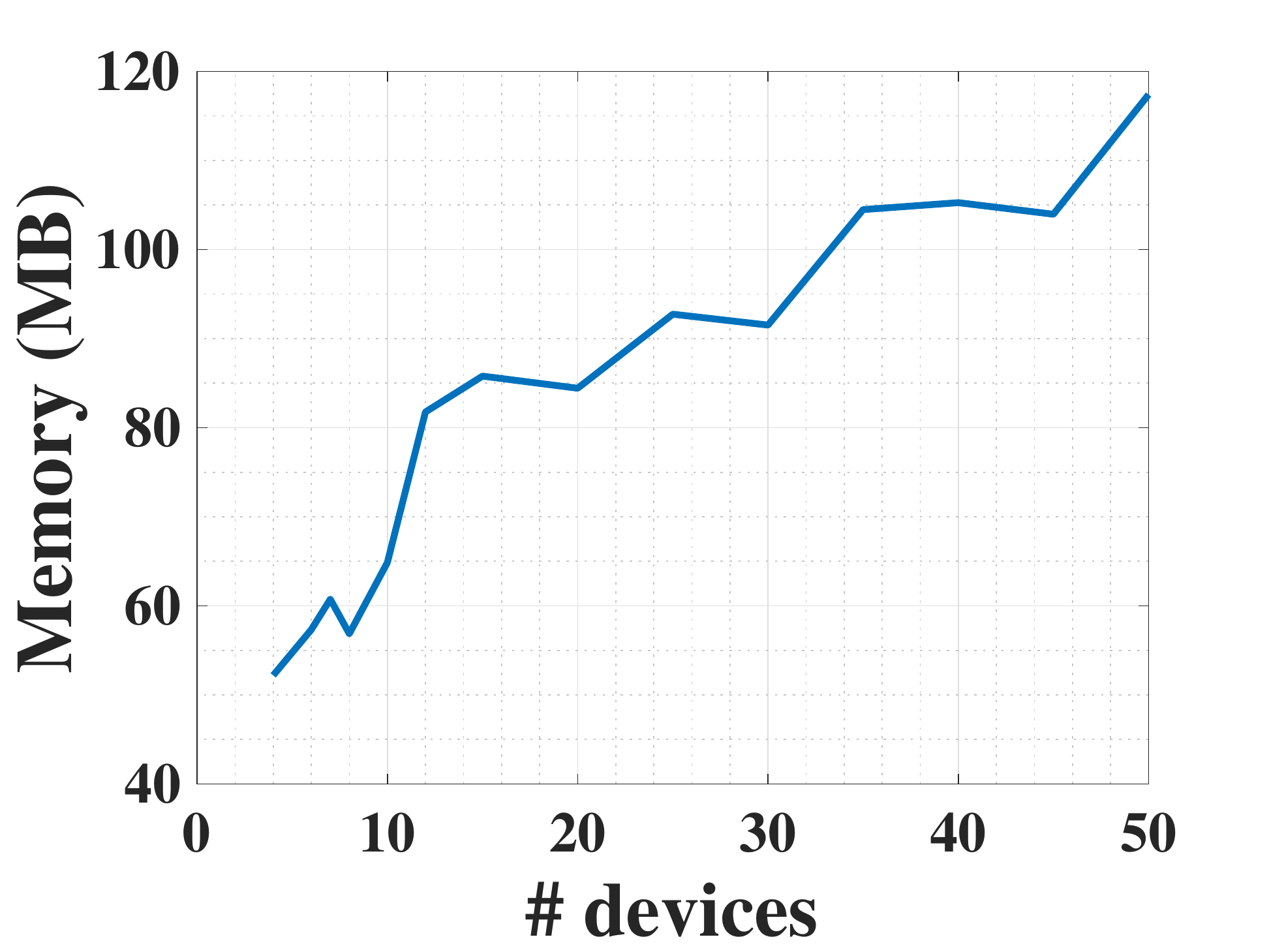}
        \caption{}
        \label{fig:memory}
    \end{subfigure}
    \caption{(a): CPU time vs IoT system size. (b): Memory usage vs IoT system size.}
    \label{fig:eval}
\end{figure}

The time and memory complexity of our framework are shown in Figure \ref{fig:eval}. From the figure we can see that, in reality, it only takes around 1.2 seconds and 120MB of memory to generate the attack graph and perform attack graph analysis for an IoT system with 50 devices. The CPU time and memory consumption grow almost linearly with the number of devices. Our graph analysis algorithms will not asymptotically increase time complexity on top of the attack graph generation algorithm because the shortest attack trace algorithm only traverses the graph once. Though the time complexity of the blast radius algorithm is bounded by the sum of the number of traces to each node, in practice, this number is at the scale of $O(n^2)$ where $n$ is the number of devices. 

%% file: sections/future_work.tex
\section{Limitations and Future Work}\label{sec:future}
Our framework uses hard-coded Prolog rules to represent direct and indirect physical dependencies between different devices. Since these rules are universal in all IoT systems, they only need to be written once and can then be copied to the Prolog rule files of all the IoT systems. And our identified six physical channels (i.e., temperature, humidity, illuminance, voice, smoke, and water) are common in IoT systems \cite{ding2018safety}. Even if new physical quantities, such as magnetic field magnitude, might be used by some special IoT systems, we can create new Prolog rules for it and insert them into all IoT systems involving such physical quantity. In the future, we plan to use machine learning to automatically extract physical channels affected/sensed by different IoT devices. We also plan to develop a vulnerability scanner for low-power IoT devices and integrate it into our \textsc{Iota} framework. 

%% file: sections/related_work.tex
\section{Related work}\label{sec:related}
\textbf{IoT security}. Existing research works on IoT security focus on different parts of IoT systems. Ding et al. \cite{ding2018safety} proposed an approach to discover potential physical interactions across applications and generate interaction chains in an IoT system. Costin et al. \cite{costin2014large} conducted a large-scale static analysis of IoT device firmware and discovered 38 previously unknown vulnerabilities. Sugawara et al. \cite{sugawara2020light} explored device sensor vulnerability and presented a new class of audio injection attacks on IoT devices' microphones by converting the audio signal to laser beams. \cite{ronen2017iot, vasserman2011vampire} explored wireless communication protocol vulnerabilities. Gu et al. \cite{gu2020iotspy} presented an approach to sniff users' privacy by analyzing the wireless traffic. \cite{nguyen2018iotsan, celik2018soteria} focused on uncovering application-level vulnerabilities using model checking techniques. Though some of the works claim to perform system-level analysis, they still just consider a subset of the core IoT components identified by our work, thus having limited capability in detecting system-level vulnerabilities. 

\textbf{Attack graph}. Automatic attack graph construction techniques are critical for system security analysis of networked systems. There has been extensive study on attack graphs for conventional computer networks. Sheyner et al. \cite{sheyner2002automated} proposed automated generation of attack graphs based on symbolic model checking. But their framework suffers from the state space explosion issue, making it difficult to model systems with hundreds of hosts. \cite{ou2006scalable, ammann2002scalable} utilized the monotonicity assumption to design attack graphs that can be generated in polynomial time. Besides, \cite{albanese2012time, durkota2015optimal} present methods to harden computer networks using attack graphs. Attack graphs are also applied to intrusion detection systems \cite{capobianco2019employing, noel2008optimal}. 

\textbf{Attack graph analyses}. Ou et al. \cite{ou2005mulval} introduce hypothetical analysis which answers the question of ``what if there are some vulnerabilities in the system?''. Sheyner et al. \cite{sheyner2002automated} propose two analyses, i.e., the minimal set of exploits to prevent so that the attackers fail to achieve their goals and the likelihood that the attacker will succeed. Ingols et al. \cite{ingols2006practical} present automatic recommendations to improve system security by identifying a bottleneck device and patching vulnerabilities to prevent attackers from accessing the bottleneck. 
Nguyen et al. \cite{nguyen2018iotsan} propose a method to attribute safety violations to either bad apps or misconfigurations. 

%% file: sections/conclusion.tex
\section{Conclusion}\label{sec:conclusion}
In this work, we design and prototype a novel framework \textsc{Iota} for automatic, system-level IoT system security analysis. \textsc{Iota} takes system configuration and CVE database as input and generates attack graphs showing all of the potential attack traces. Our framework further analyzes the attack graph by computing metrics, viz. the shortest attack trace and blast radius, to help system administrators evaluate vulnerabilities' impacts. Evaluation results show that \textsc{Iota} is both effective ($62.8\%$ of the attack traces revealed are beyond system designers' anticipation) and highly efficient (it takes less than 1.2 seconds to analyze IoT systems of 50 devices).

%% file: sections/acknowledgement.tex
\section*{Acknowledgement}
This research was sponsored by the U.S. Army Combat Capabilities Development Command Army Research Laboratory and was accomplished under Cooperative Agreement Number W911NF-13-2-0045 (ARL Cyber Security CRA). The views and conclusions contained in this document are those of the authors and should not be interpreted as representing the official policies, either expressed or implied, of the Combat Capabilities Development Command Army Research Laboratory or the U.S. Government. The U.S. Government is authorized to reproduce and distribute reprints for Government purposes notwithstanding any copyright notation here on.